\documentclass[twocolumn,showpacs,preprintnumbers,amsmath,amssymb]{revtex4}
\input psfig.sty

\usepackage{graphicx}
\usepackage{dcolumn}
\usepackage{bm}

\begin{document}

\title{Cosmic acceleration in brane cosmology}

\author{J. S. Alcaniz} \email{alcaniz@dfte.ufrn.br}

\author{N. Pires} \email{npires@dfte.ufrn.br}

\affiliation{Departamento de F\'{\i}sica, Universidade Federal do Rio Grande do Norte, 
C. P. 1641, 59072-970, Natal-RN, Brasil}

\date{\today}

\begin{abstract}

Cosmic acceleration may be the result of unknown physical processes involving either new
fields in high energy physics or modifications of gravitation theory. In the latter
case, such modifications are usually related to the existence of extra dimensions
(which is also required by unification theories), giving rise to the so-called brane
cosmology. In this paper we investigate the phenomenon of the acceleration of the
Universe in a particular class of brane scenarios in which a large scale modification
of gravity arises due to a gravitational \emph{leakage} into extra dimensions. By using
the most recent supernova observations we study the transition
(deceleration/acceleration) epoch as well as the constraints imposed on the parameters
characterizing the model. We show that these models provide a good description for the
current supernova data, which may be indicating that the existence of extra dimensions
play an important role not only in fundamental physics but also in cosmology.

\end{abstract}

\maketitle

\section{Introduction}

The current idea of a \emph{negative-pressure} dominated universe seems to be inevitable
in
light of the impressive convergence of the recent observational results (see, e.g.,
\cite{revde} for a review). This in turn has led cosmologists to hipothesize on the
possible existence of an exotic dark component that not only would
explain these experimental data but also would reconcile them with the inflationary
flatness prediction ($\Omega_{\rm{Total}} = 1$). This extra component, or rather, its
gravitational effects is thought of as the first observational piece of evidence for
new physics beyond the domain of the standard model of particle physics, a conclusion
that has given rise to many speculations on its fundamental origin \cite{alcaniz}.

Alternatively, another possible route to deal with this dark pressure problem could be a
modification in gravity instead of any adjustment to the energy content of the
Universe. This idea naturally brings to light another important question at the
interface of fundamental physics and cosmology: extra dimensions. As is well
known the existence of extra dimensions is required in various theories beyond the
standard model of particle physics, especially in theories for unifying gravity and
the other fundamental forces, such as superstring or M theories. As suggested in Ref.
\cite{add}, extra dimensions may also provide a possible explanation for the huge
difference between the two fundamental energy scales in nature, namely, the electroweak
and Planck scales [$M_{Pl}/m_{EW} \sim 10^{16}$] (see also \cite{randall}).

In the cosmological context, the role of extra spatial dimensions as source of the
dark pressure is translated into the so-called brane world (BW) cosmologies
\cite{braneS}. The general principle behind such models is that our 4-dimensional
Universe would be a surface or a brane embedded into a higher dimensional bulk
space-time on which gravity and only gravity could propagate. Brane world scenarios has
been a topic of much interest recently. In \cite{ss}, for instance, a class of BW models
which admit a
wider range of possibilities for the dark pressure than do the usual dark energy
scenarios was investigated. An interesting feature of this class of models is that the
acceleration of the Universe can be a transient phenomena, which could help reconcile
the supernova evidence for an accelerating universe with the requirements of
string/M-theory \cite{fis}. 

Another particularly interesting scenario is the one proposed by Dvali {\it{et al.}}
\cite{dgp}, which we will refer to it as DGP model. It describes a  self-accelerating
5-dimensional BW
model with a noncompact, infinite-volume extra dimension in which the whole dynamics of
gravity is governed by a competition between a 4-dimensional Ricci scalar term, induced
on the brane, and an ordinary 5-dimensional Einstein-Hilbert action. For scales below a
crossover radius $r_c$ (where the induced 4-dimensional Ricci scalar dominates), the
gravitational force experienced by two punctual sources is the usual 4-dimensional
$1/r^{2}$ force whereas for distance scales larger than $r_c$ the gravitational force
follows the 5-dimensional $1/r^{3}$ behavior. The theoretical consistency of the model,
and in particular
of its self-accelerating solution, is still a matter of debate in the
current literature (see, e.g., \cite{luty}). From the observational viewpoint, however,
DGP
models have been sucessufully tested
in many of their predictions, ranging from local gravity to cosmological observations
\cite{deff1,alc1,alc2,lue}. 

In this paper we are particularly interested in testing the viability of DGP scenarios 
in light of the latest supernova (SNe Ia) data, as provided recently by Riess {\it et
al.} \cite{rnew}. The sample used, which consists of 157 \emph{gold} events
distributed over the redshift interval $0.01 \lesssim z \lesssim 1.7$, is the
compilation
of the best observations made so far by the two supernova search teams plus 16 new
events observed by \emph{HST}. In agreement with other independent analyses, it is shown
that these models constitute a good explanation for the current SNe Ia observations and,
hence, that the presence of extra dimensions may provide a viable alternative for the
dark pressure problem.

\section{Basic equations and the transition epoch}

In DGP models, the modified Friedmann's equation due to the presence of an
infinite-volume extra dimension reads \cite{deff1}
\begin{equation} 
\left[\sqrt{\frac{\rho}{3M_{pl}^{2}} + \frac{1}{4r_{c}^{2}}} +
\frac{1}{2r_{c}}\right]^{2} = H^{2} 
+ \frac{k}{R(t)^{2}},
\end{equation} 
where $k$, $R(t)$, $H$ and $\rho$ are, respectively, the curvature
parameter of the spatial section, the cosmological scale factor, the Hubble parameter
and the energy
density of the cosmic fluid (which we will assume to be composed only of
nonrelativistic particles). The crossover scale defining the gravitational interaction
among particles located on the brane is expressed as $r_c = M_{pl}^{2}/2M_{5}^{3}$,
where $M_{pl}$ is the Planck mass and $M_5$ is the 5-dimensional reduced Planck mass.
Note that whenever the condition $\rho/3M_{pl}^{2} >> 1/r_{c}^{2}$ is valid, DGP and
standard models are analogous so that the cosmological evolution for the early
stages of the Universe (when the above condition holds) is exactly the same in both
scenarios.

Equation (1) also implies that the normalization condition is given by
\begin{equation}
\Omega_k + \left[\sqrt{\Omega_{\rm{r_c}}} + \sqrt{\Omega_{\rm{r_c}} +
\Omega_{\rm{m}}}\right]^{2} = 1
\end{equation}
where $\Omega_{\rm{m}}$ and $\Omega_k$ are, respectively, the matter and curvature
density parameters (defined in the usual way) and 
\begin{equation}
\Omega_{\rm{r_c}} = 1/4r_c^{2}H_o^{2},
\end{equation} 
is the density parameter associated to the crossover radius $r_c$. For a flat universe 
($\Omega_k = 0$), Eq. (2) reduces to $\Omega_{\rm{r_c}} = (1 -
\Omega_{\rm{m}})^{2}/4$.

\begin{figure}
\vspace{.2in}
\centerline{\psfig{figure=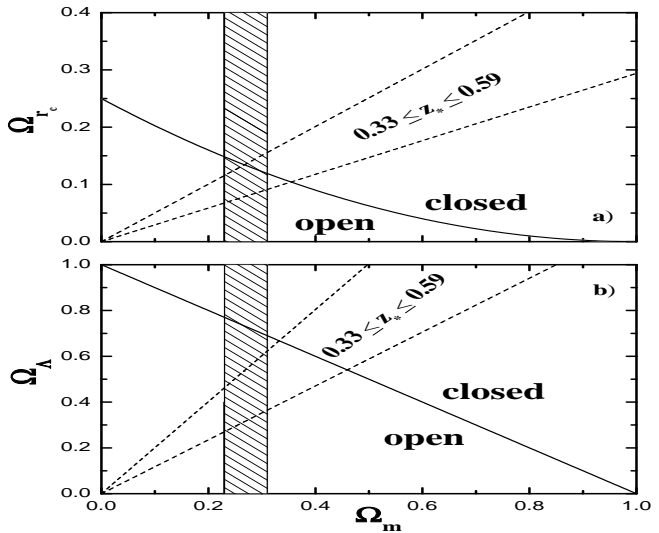,width=3.5truein,height=3.0truein,angle=0}
\hskip 0.1in}
\caption{{\bf{a)}} $\Omega_{\rm{m}} - \Omega_{\rm{r_c}}$ plane for the transition
redshift lying in the interval $0.33 \leq z_* \leq 0.59$, according to \cite{rnew}. The
vertical hachured rectangle stands for the estimate of matter density parameter as
provided by WMAP, i.e., $\Omega_{\rm{m}} = 0.27 \pm 0.04$ \cite{wmap}. {\bf{b)}} The
same as in Panel (a) for $\Lambda$CDM scenarios. Note that the coasting empty case
$\Omega_{\rm{m}} = \Omega_{\rm{r_c}} = \Omega_{\Lambda} = 0$ is always compatible.}
\end{figure}

\begin{figure*}
\vspace{.2in}
\centerline{\psfig{figure=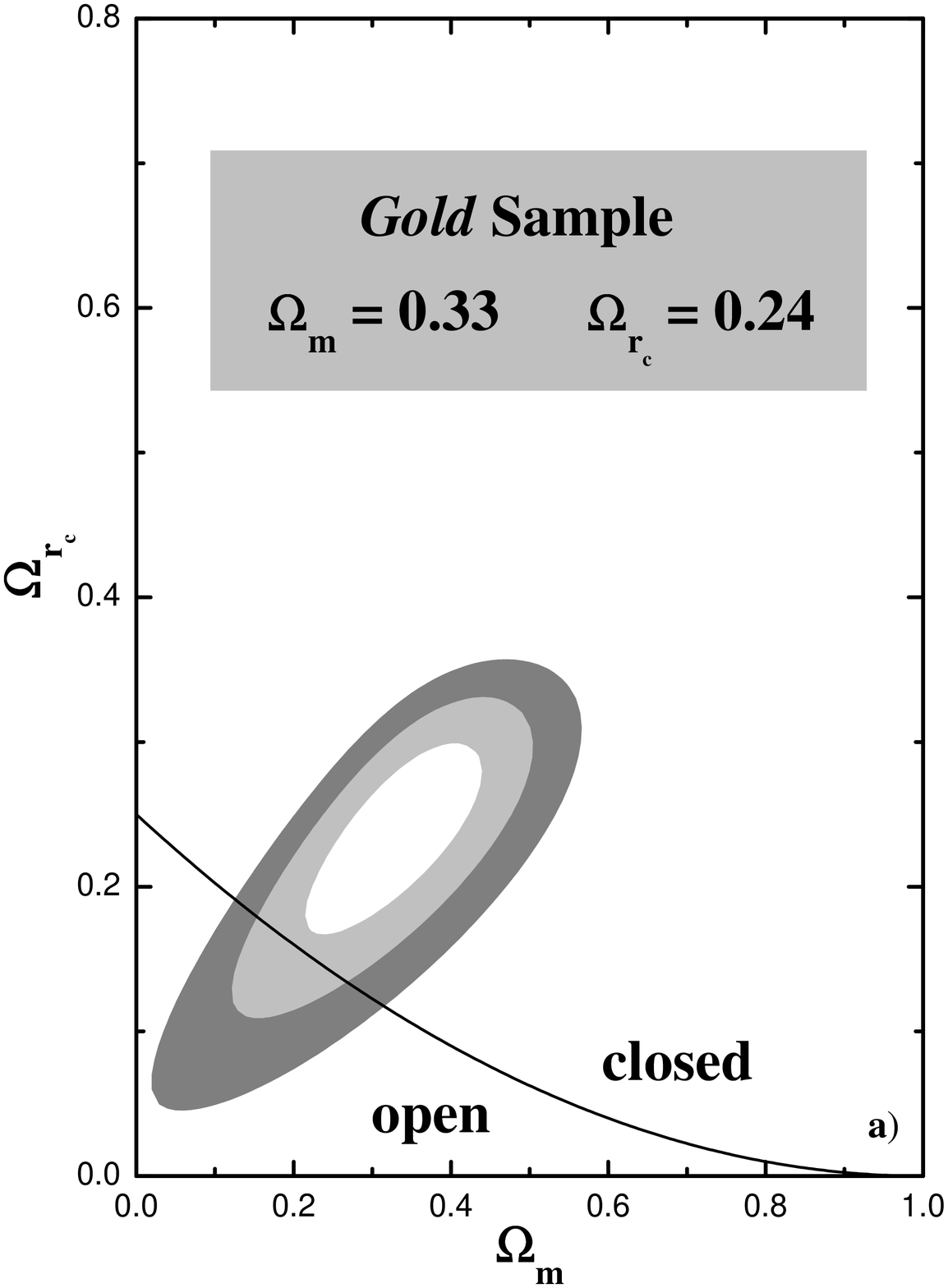,width=3.3truein,height=2.8truein,angle=0} 
\psfig{figure=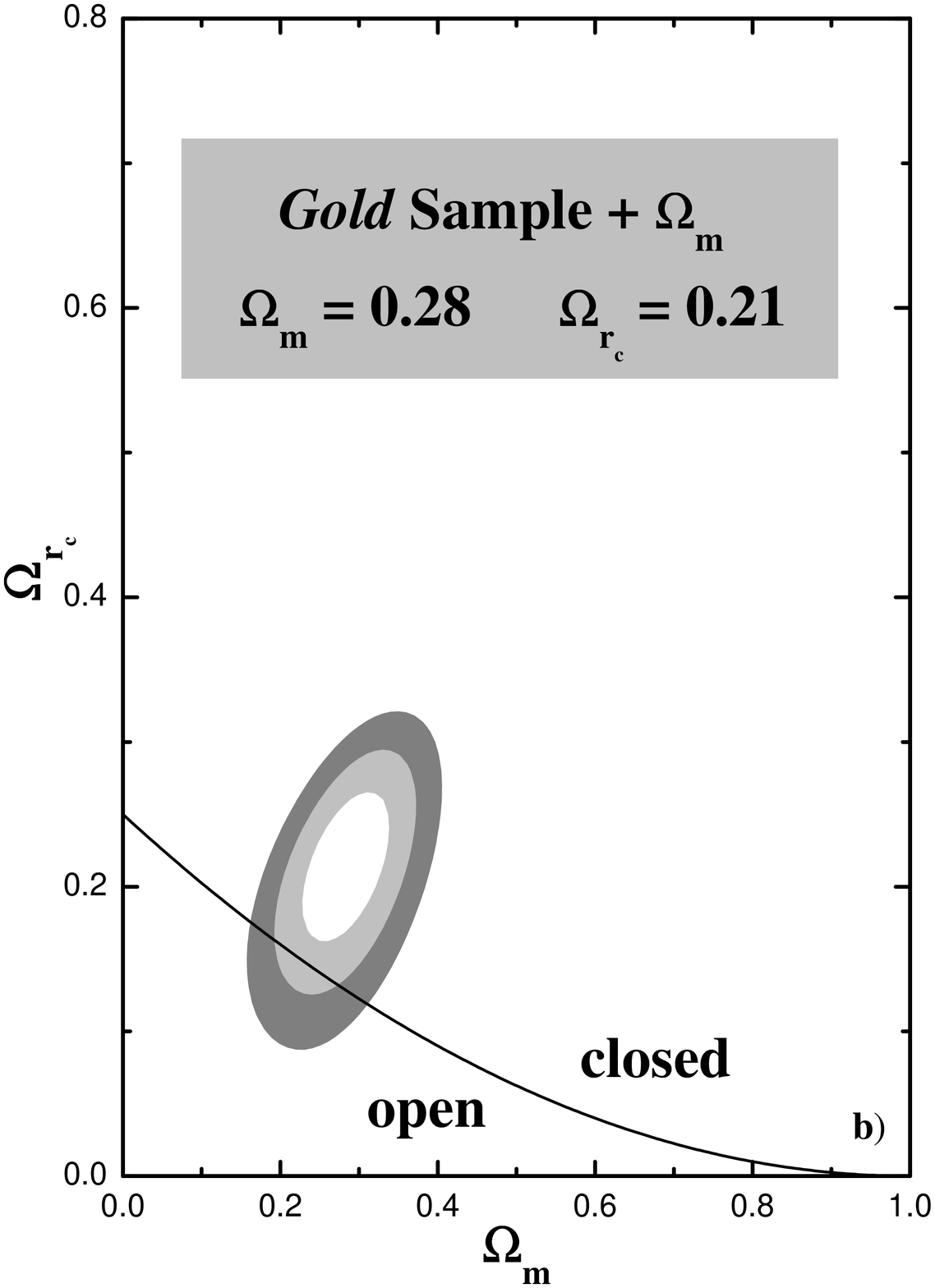,width=3.3truein,height=2.8truein,angle=0} 
\hskip 0.1in}
\caption{{\bf{a)}} Confidence regions ($68.3\%$, $95.4\%$ and $99.7\%$) in the
$\Omega_{\rm{m}} - \Omega_{\rm{r_c}}$ plane by considering the \emph{gold} sample of
Riess et al. \cite{rnew}. Note that the area corresponding to the confidence intervals
is considerably reduced when compared with previous SNe Ia analyses
\cite{deff1,avelino}. {\bf{b)}} The same as in Panel (a) by assuming a Gaussian prior
on the matter density parameter, $\Omega_{\rm{m}} = 0.27 \pm 0.04$. For this analysis
the best-fit model occurs at $\Omega_{\rm{m}} = 0.28$ and $\Omega_{\rm{r_c}} = 0.21$
with $\chi_{min}^2/\nu \simeq 1.13$.}
\end{figure*}

In order to study the acceleration phenomenon in these scenarios we derive the
deceleration parameter, defined as \cite{zhu}
\begin{equation}
q(z) = \frac{1}{2} \frac{d\mbox{ln}{\cal{F}}^2(z; \Omega_{j})}{d\mbox{ln}(1 + z)} -
1,
\end{equation}
where $j$ stands for $m$, $r_c$ and $k$ and the dimensionless function ${\cal{F}}(z;
\Omega_{j})$ is given by
\begin{equation}
{\cal{F}} = \left[\Omega_k (1 + z)^{2} + \left(\sqrt{\Omega_{\rm{r_c}}} + 
\sqrt{\Omega_{\rm{r_c}} + \Omega_{\rm{m}}(1 + z)^{3}}\right)^{2}\right]^{1/2}.
\end{equation}
From the Eqs. (4) and (5), it is possible to obtain the transition redshift $z_{*}$ at
which the Universe switches from deceleration to acceleration or, equivalently, the
redshift
at which the deceleration parameter vanishes. Regardless of the geometry adopted,
$z_{*}$ can always be expressed in a closed analytic form, i.e., 
\begin{equation}
(1 + z_*)_{q = 0} = 2\left(\frac{\Omega_{\rm{r_c}}}{\Omega_{\rm{m}}}\right)^{1/3}.
\end{equation}
If now we restrict our analysis to the flat case, in accordance to CMB measurements
\cite{wmap,beast}, and assume  $\Omega_{\rm{m}} = 0.3$ ($\Omega_{\rm{r_c}} = 0.1225$),
as suggested by clustering estimates \cite{calb}, Eq.
(6) provides $z_* \simeq 0.48$, which is surprisingly close to the transition redshift
estimated from the current SNe Ia data, i.e., $z_* = 0.46 \pm 0.13$ \cite{rnew} (see
also \cite{daly} for other recent estimates of $z_*$). For the sake of comparison, we
also
compute $z_*$ for the so-called
concordance model, namely, a flat universe with $\Omega_{\rm{m}} = 0.27$ and a vacuum
energy contribution of $\Omega_{\Lambda} = 0.73$ ($\Lambda$CDM). Such a model yields
$z_* \simeq 0.75$, which is off by $\sim 2\sigma$ from the central value estimated in
\cite{rnew}.

In Fig. 1a we show the $\Omega_{\rm{m}} - \Omega_{\rm{r_c}}$ plane for the transition
redshift lying in the interval $0.33 \leq z_* \leq 0.59$, which corresponds to $\pm
1\sigma$ of the value for $z_*$ given in \cite{rnew}. Note that a considerable
portion of this parameter space is compatible with such a constraint. The vertical
hachured rectangle stands for the estimate of matter density parameter as provided by
WMAP, i.e., $\Omega_{\rm{m}} = 0.27 \pm 0.04$ \cite{wmap}. In particular, for the
best-fit values of $\Omega_{\rm{m}}$ and $\Omega_{\rm{r_c}}$ obtained from the current
gravitational lensing data \cite{alc1} we find $z_* \simeq 0.29$ while for those
estimated from recent dating of high-$z$ objects \cite{alc2} we obtain $z_* \simeq
0.22$. For comparison, we show in Fig. 1b a similar analysis for $\Lambda$CDM scenarios
(whose expression for the transition redshift is given by $z_* =
(2\Omega_{\Lambda}/\Omega_{\rm{m}})^{1/3} - 1$). As mentioned earlier, the point
corresponding to the concordance model
($\Omega_{\rm{m}} = 0.27$, $\Omega_{\Lambda} = 0.73$) is outside the region covered by
the estimate of $z_*$ considered.

\section{statistical analysis}

In this section we test the viability of DGP scenarios through a statistical analysis
involving the most recent SNe Ia data, as provided recently by Riess {\it et al.}
\cite{rnew}. The total sample presented in \cite{rnew} consists of 186 events
distributed over the redshift interval $0.01 \lesssim z \lesssim 1.7$ and constitutes
the compilation of
the best observations made so far by the two supernova search teams plus 16 new events
observed by \emph{HST}. This total data-set was divided into ``high-confidence''
(\emph{gold}) and ``likely but not certain'' (\emph{silver}) subsets.  Here, we will
consider only the 157 events that constitute the so-called \emph{gold} sample. In what
follows we briefly outline
our main assumptions for this analysis (see \cite{sne} for some recent SNe Ia analyses).

The predicted distance modulus for a supernova at redshift $z$, given a set of
parameters $\mathbf{s}$, is
\begin{equation}
\mu_p(z|\mathbf{s}) = m - M = 5\mbox{log} d_L + 25,
\end{equation}
where $m$ and $M$ are, respectively, the apparent and absolute magnitudes, the complete
set of parameters is $\mathbf{s} \equiv (H_o, \Omega_{j})$ and $d_L$ stands for the
luminosity distance (in units of megaparsecs),
\begin{equation}
d_L = \frac{c(1 + z)}{H_o\sqrt{|\Omega_k|}}S_k\left[\sqrt{|\Omega_k|}\int_{x'}^{1} {dx
\over x^{2}{\cal{F}}(x; \Omega_{j})}\right],
\end{equation}
with $x' = {R(t) \over R_o} = (1 + z)^{-1}$ being a convenient integration variable,
${\cal{F}}(x; \Omega_{j})$ the expression given by Eq. (6) and  $S_k$ a function
defined by one of the following forms: $S_k(r) \equiv  \mbox{sinh}(r)$, $r$, and
$\mbox{sin}(r)$, respectively, for open, flat and closed geometries.

We estimated the best fit to the set of parameters $\mathbf{s}$ by using a $\chi^{2}$
statistics, with
\begin{equation}
\chi^{2} = \sum_{i=1}^{157}{\frac{\left[\mu_p^{i}(z|\mathbf{s}) -
\mu_o^{i}(z|\mathbf{s})\right]^{2}}{\sigma_i^{2}}},
\end{equation}
where $\mu_p^{i}(z|\mathbf{s})$ is given by Eq. (7), $\mu_o^{i}(z|\mathbf{s})$ is the
extinction corrected distance modulus for a given SNe Ia at $z_i$, and $\sigma_i$ is
the uncertainty in the individual distance moduli, which includes uncertainties in
galaxy redshift due to a peculiar velocity of 400 km/s. The Hubble parameter $H_o$ is
considered a``nuisance" parameter so that we marginalize over it by using the analytical
method of Ref. \cite{wang}.

Figure 2a shows the confidence regions ($68.3\%$, $95.4\%$ and $99.7\%$) in the
$\Omega_{\rm{m}} - \Omega_{\rm{r_c}}$ plane by considering the \emph{gold} sample of
Riess et al. \cite{rnew}. Compared to Fig. 1 of \cite{avelino} and Fig. 2 of
\cite{deff1} (which used the then available SNe Ia data), the \emph{gold} sample
studied here reduces considerably the area corresponding to the confidence intervals.
The best-fit parameters for this analysis are $\Omega_{\rm{m}} = 0.33$ and
$\Omega_{\rm{r_c}} = 0.24$ with $\chi_{min}^2/\nu \simeq 1.12$ ($\nu \equiv$ degrees of
freedom). At 95\% c.l. we obtain $0.24 \leq \Omega_{\rm{m}} \leq 0.43$ and $0.17 \leq
\Omega_{\rm{r_c}} \leq 0.29$. In particular, the above relative value of $\chi^2$ is
slightly smaller than the one we found for the concordance scenario,
$\chi_{min}^2/\nu \simeq 1.14$, and equal to the value we obtained for the $\Lambda$CDM
with arbitrary curvature. By restricting the analysis to the flat case, we note that the
data favour a lower value of the matter density parameter, i.e., $\Omega_{\rm{m}} =
0.21$ ($\Omega_{\rm{r_c}} = 0.156$) with  $\chi_{min}^2/\nu \simeq
1.13$. Such a value, however, is inside the 2$\sigma$ interval of the current WMAP
estimates of the quantity of matter in the Universe,
$\Omega_{\rm{m}} = 0.27 \pm 0.04$ \cite{wmap}, and therefore does not imply any possible
conflit between the predictions of the model and the independent measurements of
$\Omega_{\rm{m}}$ (see \cite{avelino,deffC} for a discussion on this point).

By using SNe Ia-independent constraints in the $\Omega_{\rm{m}} - \Omega_{\rm{r_c}}$
plane yields more precise limits on these parameters due to the degeneracy between them
for SNe Ia data. Thus, in Fig. 2b we show a similar analysis to Fig. 2a having assumed
a Gaussian prior on the matter
density parameter,
$\Omega_{\rm{m}} = 0.27 \pm 0.04$, as provided by WMAP team \cite{wmap}. The parameter
space now is considerably reduced relative to the former analysis, with the best-fit
model occurring at $\Omega_{\rm{m}} = 0.28$ and $\Omega_{\rm{r_c}} = 0.21$
($\chi_{min}^2/\nu \simeq 1.13$). Such a model corresponds to an accelerating universe
with $q_o \simeq -0.65$ and a total expanding age of $t_o \simeq 9.9h^{-1}$ Gyr. Note
that the best-fit value for $\Omega_{\rm{r_c}}$ leads to an estimate of the crossover
scale $r_c$ in terms of the Hubble radius $H_o^{-1}$ (see Eq. 3),
\begin{equation}
r_c \simeq 1.09 H_o^{-1},
\end{equation} 
which is slightly smaller than the one obtained in Ref. \cite{deff1} by using the old
SNe Ia data (see also \cite{alc2} for summary of the current estimates of $r_c$). By
fixing $\Omega_k = 0$, we find $\Omega_{\rm{m}} = 0.23$ ($\Omega_{\rm{r_c}} = 0.148$)
with  $\chi_{min}^2/\nu \simeq 1.15$. This particular value of  $\Omega_{\rm{r_c}}$
corresponds to a crossover distance between 4-dimensional and 5-dimensional gravities
of the order of $r_c \simeq 1.3 H_o^{-1}$. In Table I we summarize the main results of
the paper.

\begin{table}
\caption{Best-fit values for $\Omega_{\rm{m}}$, $\Omega_{\rm{r_c}}$ and $r_{c}$}
\begin{ruledtabular}
\begin{tabular}{lcrl}
Test& $\Omega_{\rm{m}}$ (flat) & $\Omega_{\rm{r_c}}$ (flat) &$r_{c}$\footnote{in units
of $H_o^{-1}$} (flat)\\
\hline \hline \\
SNe Ia & 0.33 (0.21) & 0.24 (0.156) & 1.02 (1.26)\\
SNe Ia + $\Omega_{\rm{m}}$ & 0.28 (0.23) & 0.21 (0.148) & 1.09 (1.30)\\
\end{tabular}
\end{ruledtabular}
\end{table}

\section{Conclusion}

The results of observational cosmology in the past years have opened up an unprecedented   
opportunity to establish a more solid connection between Fundamental Physics and
Cosmology. Surely, the most remarkable finding among these results comes from SNe Ia
observations which suggest that the cosmic expansion is undergoing a late time
acceleration. Such a phenomenon has been often explained from two different ways, i.e.,
either by considering the presence of a negative-pressure dark component, the so-called
dark energy, or by assuming the existence of extra spatial dimensions, as motivated by
recent developments in particle physics. In this paper we have followed the second
route. We have studied the transition (deceleration/acceleration) epoch in the context
of DGP models and shown that these scenarios based on a large 
scale modification of gravity are in good agreement with the current SNe Ia data for
values of $\Omega_{\rm{m}} = 0.33^{+0.10}_{-0.09}$  and $\Omega_{r_{c}} =
0.24^{+0.05}_{-0.07}$ (95.4\% c.l.). By
assuming a SNe Ia-independent constraint on the matter density parameter, we have
found $\Omega_{\rm{r_c}} = 0.21$, leading to an estimate of the crossover scale $r_c
\simeq 1.09 H_o^{-1}$. If flatness is imposed, then we find
$\Omega_{\rm{m}} = 0.23$ ($\Omega_{\rm{r_c}} = 0.148$), which corresponds to a
acceleration universe with $q_o \simeq -0.65$ and a total expanding age of $t_o \simeq
9.5h^{-1}$ Gyr. 

In summary, what we have shown is that at least at the level of
background tests (like tests involving SNe Ia measurements), DGP models constitute a
viable alternative for the dark energy or dark pressure problem. This may be understood
as an 
indication that the existence of extra dimensions play an important role not only in
fundamental physics but also in cosmology.

\begin{acknowledgments}
The authors are very grateful to A. G. Riess for valuable discussions and to G. S.
Fran\c{c}a for a critical reading of the manuscript. JSA is supported by CNPq
(305205/02-1) and CNPq (62.0053/01-1-PADCT III/Milenio). NP is supported by
PRONEX/CNPq/FAPERN.
\end{acknowledgments}


\end{document}